\documentclass[aps,prb,twocolumn,10pt]{revtex4-1}
\usepackage{psfrag,graphics,graphicx}
\usepackage{amsmath}
\usepackage[OT1]{fontenc}

\newcommand{\ms}[1]{\mbox{\scriptsize{#1}}}

\def\bea#1\eea{\begin{align}#1\end{align}}
\def\be#1\ee{\begin{align}#1\end{align}}

\def\bean#1\eean{\begin{align*}#1\end{align*}}

\def\D{\Delta}
\def\e{\epsilon}
\def\o{\omega}

\def\l{\lambda}
\def\L{\Lambda}
\def\d{\delta}

\def\s{\sigma}
\def\m{\mu}
\def\n{\nu}
\def\ra{\rightarrow}

\def\bd{b^{\dagger}}
\def\ra{\rightarrow}
\def\nn{\nonumber}

\def\ben{\begin{enumerate}}
\def\een{\end{enumerate}}

\newcommand{\bra}[1]{\left< #1 \right|}
\newcommand{\ket}[1]{\left| #1 \right>}

\begin{document} 
\title{Order parameter, correlation functions and fidelity susceptibility for the BCS model in the thermodynamic limit}
\author{Omar El Araby$^{1,2}$}
\author{Dionys Baeriswyl$^{1,3}$}
\affiliation{$^1$~Department of Physics, University of Fribourg, Chemin du mus\'ee 3, CH-1700 Fribourg\\$^2$~Institute for Theoretical Physics, University of Amsterdam, Science Park 904, NL-1090 Amsterdam\\$^3$~International Institute of Physics, 59078-400 Natal-RN, Brazil}
\begin{abstract}
The exact ground state of the reduced BCS Hamiltonian is investigated numerically  for large system sizes and compared with the BCS ansatz. A ``canonical'' order parameter is found to be equal to the largest eigenvalue of Yang's reduced density matrix in the thermodynamic limit. Moreover, the limiting values of the exact analysis agree with those obtained for the BCS ground state. Exact results for the ground state energy, level occupations and a pseudospin-pseudospin correlation function are also found to converge to the BCS values already for relatively small system sizes. However, discrepancies persist for a pair-pair correlation function, for inter-level correlations of occupancies and for the fidelity susceptibility, even for large system sizes where these quantities have visibly converged to well-defined limits. Our results indicate that there exist non-perturbative corrections to the BCS predictions in the thermodynamic limit.
\end{abstract}

\maketitle

\section{Introduction}
\label{sec1}
The microscopic theory of Bardeen, Cooper and Schrieffer (BCS) \cite{bcs1} represents arguably the central paradigm of superconductivity, but it plays also a crucial role for superfluid helium-3 
\cite{leggett}, ultracold gases of fermionic atoms \cite{giorgini, bloch}, atomic nuclei \cite{broglia} and neutron stars \cite{wilczek}. The theory involves two elements, on the one hand the so-called reduced BCS Hamiltonian, where only scattering processes between zero-momentum pairs of fermions are taken into account, on the other hand a variational ansatz for the ground state of this Hamiltonian, a coherent superposition of products of pair wave functions. In this paper we address the question whether the BCS ansatz is the exact ground state of the reduced BCS Hamiltonian in the limit of an infinitely large system size. An early argument for the asymptotic validity of the BCS ansatz was given by Anderson \cite{anderson}, who pointed out that BCS theory should be ``nearly valid'' because in the limit of large numbers ``quantum fluctuations die out''. Later explicit calculations showed that indeed the ground state energy, level occupation and the free energy were exactly predicted by BCS theory in the thermodynamic limit \cite{bogoliubov, muehlschlegel, bursill}. Moreover, for a specific single-particle spectrum (``step model'') Mattis and Lieb  concluded that the BCS wave function was exact in this limit \cite{mattislieb}. 

Our results, based on Richardson's exact solution of the reduced BCS Hamiltonian  
\cite{richardson1, richardson2}, confirm that many quantities, for instance level occupancies or the ground state energy, are predicted accurately by BCS theory in the thermodynamic limit. This is also true for an order parameter, defined according to Yang's concept of off-diagonal long-range order (ODLRO) \cite{yangodlro}. However, for other quantities, such as a pair-pair correlation function, inter-level occupancy fluctuations and the fidelity susceptibility, BCS predictions are found to differ from the (numerically) exact results, even for very large system sizes. 

The paper is organized as follows. Section \ref{sec2} describes Richardson's exact solution for the eigenstates of a simplified form of the reduced BCS Hamiltonian. Section \ref{sec3} deals with the ground state, on the one hand in BCS approximation, on the other hand by evaluating the exact solution numerically. The exact ground state energy is shown to approach rapidly the BCS prediction as a function of system size. In Section \ref{sec4} it is shown that Yang's ODLRO is encoded in a ``canonical'' order parameter, which is found to converge to the BCS result in the thermodynamic limit. Correlation functions involving HOMO and LUMO orbitals are calculated in Section \ref{sec5}. While for pseudo spin operators BCS theory is again found to agree with the $L\rightarrow\infty$ limit of the exact solution, this is not true for pair operators nor for level occupancies. A similar discrepancy is found for the ground state fidelity susceptibility, as shown in Section \ref{sec6}. The results are summarized in Section \ref{sec7}.

\section{Hamiltonian and its exact eigenstates}
\label{sec2}

The reduced BCS Hamiltonian in the form introduced by Richardson \cite{richardson1} for describing nucleons coupled by pairing forces is
\bea
H=\sum_{\nu\sigma}\varepsilon_{\nu}c_{\nu\sigma}^\dag c_{\nu\sigma}-\frac{g}{L}\sum_{\mu,\nu,\mu\ne\nu}
c_{\mu\uparrow}^\dag c_{\mu\downarrow}^\dag c_{\nu\downarrow}c_{\nu\uparrow},
\label{reddbcs}
\eea
where $c^{\dagger}_{\nu\sigma}$ and $c_{\nu\sigma}$ are, respectively, creation and annihilation operators for fermions in level $\nu$ with spin $\sigma$ and $\varepsilon_{\nu}=-\frac{W}{2}+\frac{W}{2L}(2\nu-1)$, $\nu=1,\dots,L$. We use the width $W$ as unit of energy, $W=1$, and of course assume $g>0$. The Hamiltonian has particle-hole symmetry, and therefore the chemical potential vanishes if the number of fermions equals $L$ (half filling), the case considered in this paper.
The calculations presented below can readily be performed for other forms of the single-particle spectrum, for instance for the tight-binding spectrum of the square lattice, but to discuss the generic large $L$ behavior it is advantageous to choose a spectrum exhibiting neither degeneracies nor van Hove singularities.  

In the reduced BCS Hamiltonian (\ref{reddbcs}) all levels are coupled equally, i.e. the interaction has infinite range for $L\ra\infty$ in the space of quantum numbers $\n$ (in $\vec{k}$-space for Bloch electrons). In classical statistical mechanics infinitely long-range interactions are generally believed to be treated exactly by mean-field theory. This suggests that the mean-field description of BCS for the Hamiltonian (\ref{reddbcs}) is also exact in the thermodynamic limit. There is however a loophole in this argument. A quantum system in $d$ dimensions corresponds to a classical system in $d+1$ dimensions. On the additional axis representing time the interaction does not have to be long-ranged. Therefore it is worthwhile to investigate the large $L$ limit of Richardson's exact solution in detail.

The eigenstates of the Hamiltonian (\ref{reddbcs}) can be classified according to the number of singly-occupied levels. The ground state belongs to the subspace where all levels are either doubly occupied or empty ($L$ even). Within this subspace the operators $c_{\n\s}^\dag c_{\n\s}$ are identical to $\bd_\n b_\n$, where 
\bea
b_\nu^\dag=c_{\nu\uparrow}^\dag c_{\nu\downarrow}^\dag \mbox{ and } 
b_\nu=c_{\nu\downarrow}c_{\nu\uparrow}
\eea 
create and annihilate pairs, respectively. Therefore the level occupancy can be written as
\bea
n_\nu:=\sum_\sigma c_{\nu\sigma}^\dag c_{\nu\sigma}=2\bd_\n b_\n
\eea
and the Hamiltonian (\ref{reddbcs}) is equivalent to 
\bea
H=2\sum_{\nu}\varepsilon_{\nu}b_{\nu}^{\dag}b_{\nu}-\frac{g}{L}\sum_{\mu,\nu,\mu\ne\nu}b_\mu^\dag b_\nu
\label{redbcs1}
\eea
in the subspace where single occupancy is forbidden.

The operators $b_\n,\bd_\n$ and $n_\n$ can be combined to pseudospin operators $\vec{s}_\n$ \cite{anderson} with components
\bea
s_{\n x}=\frac{1}{2}(b_\n+\bd_\n)\nn\\
s_{\n y}=\frac{i}{2}(b_\n-\bd_\n)\nn\\
s_{\n z}=\frac{1}{2}(n_\n-1).
\label{eq:pseudospin}
\eea
In terms of these operators the Hamiltonian (\ref{redbcs1}) reads
\bea
H=4\sum_{\n}\e_\n s_{\n z}-\frac{g}{L}\sum_{\substack{\m,\n \\ \m\ne\n}}s_{\m +}s_{\n -},
\label{redbcs2}
\eea
where $s_{\m \pm}=s_{\n x}\pm i s_{\n y}$, and represents an XY ferromagnet with long-range interaction in an inhomogeneous transverse field. This Hamiltonian is part of a larger family of integrable models, for which eigenstates and eigenvalues were found by Gaudin \cite{gaudindiago}. Integrability means that there exist $L$ operators $R_\n$, $\nu=1,...,L$, which commute among themselves and with the Hamiltonian. For our model the $R$-operators are \cite{cambiaggio}
\bea
R_\n=s_{\n z}+\frac{g}{L}\sum_{\m,\m\ne\n}\frac{\vec{s}_\m\cdot\vec{s}_\n}{\e_\m-\e_\n}.
\label{eq:rop}
\eea
One readily verifies that for the case considered here ($N=L$) the Hamiltonian (\ref{redbcs2}) can be written as 
\bea
H=2\sum_{\n=1}^L\e_\n R_\n,
\eea
which therefore also commutes with all operators $R_\n$.

The exact eigenstates of the Hamiltonian (\ref{reddbcs}) for $M=\frac{L}{2}$ pairs have the form 
\cite{richardson1, richardson2}
\bea
\vert\Psi_R\rangle=\prod_{i=1}^M B_i^{\dagger}\vert 0\rangle,\quad B_i^{\dagger}=\sum_{\nu=1}^L\frac{1}{2\varepsilon_{\nu}-\lambda_i}b_{\nu}^{\dagger},
\label{rich}
\eea
where $\ket{0}$ is the vacuum state, $c_{\n\s}\ket{0}=0$, and the ``rapidities'' $\lambda_i$ satisfy the Richardson (or Bethe) equations 
\bea
1-\frac{g}{L}\sum_{\nu=1}^L\frac{1}{2\varepsilon_{\nu}-\lambda_k}-\frac{g}{L}\sum_{i,i\ne k}^M\frac{2}{\lambda_k-\lambda_i}=0.
\label{betheeq}
\eea
The systems for which these equations can be directly solved are rather small, but recent algorithmic progress \cite{solver} allows us to study much larger sizes $L$ than before.
Analytical insight has been provided by Gaudin \cite{gaudinmodexres} in the continuum limit ($L\rightarrow\infty$), using an analogy to electrostatics. His result was used to show 
\cite{richardson3, sierra} that the BCS equations for the gap, the chemical potential and the ground state energy are reproduced in the thermodynamic limit. The low energy excitations have also been obtained by solving Richardson's equations analytically in the strong coupling limit \cite{emil}.

\section{Ground state and ground state energy}
\label{sec3}

The conventional BCS ground state is defined as
\bea
\vert\Psi_{\ms{BCS}}\rangle=\prod_{\nu}(u_{\nu}+v_{\nu}b_\nu^\dag)\vert 0\rangle,
\label{eq:grst_bcs}
\eea
where 
\bea
u_\nu=\sqrt{\frac{E_\nu+\varepsilon_\nu}{2E_\nu}}\, ,\quad
v_\nu=\sqrt{\frac{E_\nu-\varepsilon_\nu}{2E_\nu}}\, ,
\eea 
with $E_\nu=(\varepsilon_\nu^2+\Delta^2)^{1/2}$. The gap parameter $\Delta$ is determined by minimizing the energy expectation value. For the present model we obtain  
\bea
\Delta=\left(2\sinh\frac{1}{g}\right)^{-1}
\eea 
in the limit $L\rightarrow\infty$. The BCS state can also be written as 
\bea
\vert\Psi_{\mbox{\scriptsize BCS}}\rangle\propto e^{B^{\dagger}}\vert 0\rangle,\quad B^{\dagger}=\sum_{\nu}\frac{v_{\nu}}{u_{\nu}}b_{\nu}^{\dagger}. 
\label{bcs_conv}
\eea
Its projection on a subspace with a definitive number of $M$ pairs,
$\vert\Psi_{\mbox{\scriptsize BCS}}^{(M)}\rangle\propto (B^{\dagger})^M\vert 0\rangle$,
resembles the Richardson solution (\ref{rich}), but, as emphasized by Combescot and collaborators \cite{combescot}, in the BCS state all pair operators are equal ($B^\dag$), while  they are all different in the Richardson solution ($B_i^\dag$). In the thermodynamic limit the conventional and number-projected BCS ground states are expected to be equivalent, but for finite $L$ they differ,
especially for weak couplings. Thus conventional BCS theory predicts a phase transition at 
a critical coupling strength $g_c(L)$, below which the gap parameter vanishes, while there exists only a crossover for the number-projected BCS ground state \cite{braun, dukelsky}. The behavior is completely smooth for the exact solution \cite{dukelsky}. Here we concentrate on the 
region $g>g_c(L)$, where we can expect the different ground states to merge. For large $L$ the critical coupling strength is approximately given by $g_c(L)\approx [\log{(2L)}+\gamma]^{-1}$, where $\gamma$ is Euler's constant ($\gamma=0.5772157...$).

\begin{figure}[ht]
\includegraphics[width=0.48\textwidth]{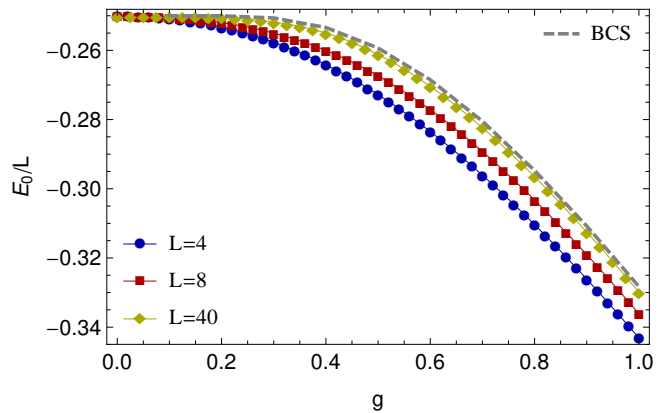}
  \caption{Ground state energy of the reduced BCS Hamiltonian. Symbols represent the exact result for different system sizes while the dashed line stands for the BCS result in the thermodynamic limit.}
\label{fig:fig1}
\end{figure}

The exact ground state can be analyzed numerically by scanning the Richardson equations from $g=0$ up to some finite value $g>0$. The computations are greatly simplified by introducing the variables 
\bea
\Lambda_\nu=\frac{g}{L}\sum_{i}\frac{1}{2\epsilon_\nu-\lambda_i},
\label{eq:substituted}
\eea
which satisfy the ``substituted Bethe equations'' \cite{solver}
\bea
\L^2_{\n}-\L_{\n}-\frac{g}{2L}\sum_{\mu,\mu\ne \nu}\frac{\L_{\mu}-\L_{\nu}}{\e_{\mu}-\e_{\nu}}=0.
\eea 
These quadratic equations can be readily solved for much larger system sizes than the original Richardson equations (\ref{betheeq}). Some quantities are simple functions of $\L_\n$. Thus the ground state energy $E_0(g)$ is given by the formula
\bea
E_0=\sum_{i=1}^M\l_i+\frac{g}{2}=\sum_{\nu=1}^L2\e_{\nu}\L_{\nu}-\frac{gM}{2}.
\eea
Results for different sizes are shown in Fig. \ref{fig:fig1}. As expected, the curves converge very rapidly towards the asymptotic limit of BCS theory,
\bea
E_0^{\ms{BCS}}\ra-\frac{L}{4}\coth\frac{1}{g}, ~~~~L\ra\infty.
\eea

\section{Order parameter}
\label{sec4}
  
The order parameter of conventional BCS theory \cite{gorkov}, 
\bea
F=\sum_{\n}\langle\Psi_0\vert b^\dagger_\n\vert\Psi_0\rangle=\frac{L\D}{g}\, ,
\label{eq:gorkov}
\eea
vanishes for a definitive number of particles and one has to search for alternatives. A ``canonical pairing parameter" has been proposed by von Delft {\it et al.} \cite{vonDelft} and adopted in other studies \cite{Tian,alexcorr},
\bea
\Phi_{\ms{can}}=\sum_{\n}\left(\langle b_{\n}^\dagger b_{\n}\rangle-
\langle c_{\nu\uparrow}^\dag c_{\nu\uparrow}\rangle
\langle c_{\nu\downarrow}^\dag c_{\n\downarrow}\rangle\right)^{\frac{1}{2}},
\eea
where we have used the notation $\langle O\rangle:=\langle\Psi_0\vert O\vert\Psi_0\rangle$. Numerical calculations for the exact ground state \cite{dukelsky} indicate that 
$\Phi_{\ms{can}}\rightarrow F$ for $L\rightarrow\infty$.
However, $\Phi_{\ms{can}}$ does not probe phase coherence and therefore the quantity
\bea
\Psi_{\ms{can}}=\sum_{\n}\left(\langle b_{\n}^\dagger b_{\n}\rangle-
\langle c_{\nu\uparrow}^\dag c_{\nu\uparrow}\rangle
\langle c_{\nu\downarrow}^\dag c_{\n\downarrow}\rangle\right)
\label{op}
\eea
was judged to be more adequate \cite{vondelft1}. Within BCS theory one has
\bea
\Psi_{\ms{can}}^{\ms{BCS}}=\sum_{\n=1}^L\left(\frac{\D}{2E_\n}\right)^2\ra\frac{L\D}{2}\arctan\frac{1}{2\D}
\mbox{ for } L\ra\infty.
\eea
This expression reaches a finite limiting value $(1/4)$ for $g\ra\infty$, while both $F$ and $\Phi_{\ms{can}}$ increase indefinitely with $g$ and represent asymptotically a pair binding energy rather than a measure of order.

The pseudospin operators (\ref{eq:pseudospin}) can be used to rewrite the order parameter 
$\Psi_{\ms{can}}$. First we notice the general relation
\bea
b_\nu^\dag b_\nu=\frac{2}{3}\vec{s}^{\ 2}_\nu+\frac{1}{2}(1-n_\nu)\, .
\eea
It is easy to see that both the BCS ansatz (\ref{eq:grst_bcs}) and the exact ground state (\ref{rich}) are eigenstates of $\vec{s}^{\ 2}_\nu$ with eigenvalue $\frac{3}{4}$ and we may write
\bea
\Psi_{\ms{can}}=\frac{1}{4L}\sum_\nu \langle n_\nu\rangle(2-\langle n_\nu\rangle)\, .
\eea
$\Psi_{\ms{can}} =0$ for the filled Fermi sea, where $\langle n_\nu\rangle$ vanishes for 
$\varepsilon_\nu>\varepsilon_F$ and is equal to 2 for $\varepsilon_\nu<\varepsilon_F$. Therefore this order parameter measures deviations from the level distribution of the filled Fermi sea. This is very satisfactory, at the same time there exist other Fermi surface instabilities leading to similar level redistributions as superconductivity. Hence $\Psi_{\ms{can}}$ is somewhat less specific than Gorkov's order parameter $F$, which is based on the breaking of gauge symmetry. 

In another proposal, inspired by Yang's ODLRO, the correlation functions
\bea
C_{\m\n}=\bra{\Psi_0}b_\m^\dagger b_{\n}\ket{\Psi_0}
\label{eq:dens_matr}
\eea
are summed up to yield the parameter \cite{Tian}
\bea
\Phi_{\ms{OD}}=\sum_{\m,\n}C_{\m\n},
\eea
which is equal to $F^2$ for the BCS ground state in the limit $L\ra\infty$. However, Yang's concept of ODLRO is based on the largest eigenvalue of the matrix $C$ rather than on the sum of its matrix elements. Thus ODLRO exists if (and only if) $C$ has an eigenvalue of the order of the particle number, i.e. of the order $L$ in the present case. This is indeed true for the conventional BCS ground state, for which the correlation functions are given by (half filling)
\bea
C_{\m\n}=\begin{cases}\frac{E_\n-\e_\n}{2E_\n} ,~~\m=\n\\
\frac{\D^2}{4E_\m E_\n},~~\m\ne\n.
\end{cases}
\eea
To find the eigenvalues of the matrix $C$ we have to calculate the determinant $C-\o I$ where $I$ is the unit matrix. Introducing the quantities 
\bea
f_\n=\frac{\D}{2E_\n},~~~~g_\n=\frac{E_\n-\e_\n}{2E_\n}
\eea
we can write $C-\o I=FAF$ where $F$ is diagonal with $F_{\n\n}=f_\n$ and
\bea
A_{\m\n}=\begin{cases}a_\n,~~~~\m=\n,\\ 1,~~~~\m\ne\n,\end{cases}
\eea
with $a_\n=(g_\n-\o)/f^2_\n$. Thus the eigenvalues of $C$ are given by the zeroes of 
\bea
\det A =\prod_{\n=1}^L(a_\n-1)\left[1+\sum_{\m=1}^L\frac{1}{a_\m-1}\right].
\eea
Together with
\bea
a_\n-1=\frac{1}{\D^2}\left[(E_\n-\e_\n)^2-4E_\n^2\o\right]
\eea
we arrive at the eigenvalue equation
\bea
1+\sum_{\n=1}^L\frac{\D^2}{(E_\n-\e_\n)^2-4E_\n^2\o}=0.
\label{eigeq}
\eea
For eigenvalues $\o$ of order 1 the summand has to change sign somewhere between $\n=1$ and $\n=L$. In turn, if all the terms in the sum are negative, the eigenvalue has to be of order $L$, for which in the limit $L\ra\infty$ Eq. (\ref{eigeq}) implies
\bea
\o_{\ms{max}}^{\ms{BCS}}=\sum_{\n=1}^L\left(\frac{\D}{2E_\n}\right)^2=\Psi_{\ms{can}}^{\ms{BCS}}.
\eea
To find out whether this remarkable equality of order parameter and largest eigenvalue of the reduced density matrix remains valid beyond BCS theory, we have 
calculated both $\o_{\ms{max}}$ and $\Psi_{\ms{can}}$ for the exact ground state. The matrix elements $C_{\m\n}$ can be expressed as sums of certain determinants which depend explicitly on the rapidities $\l_i$ \cite{alexcorr} and are not simple functions of the quantities $\L_\n$. Nevertheless it turns out to be advantageous to solve first the quadratic equations for $\L_\n$ and then use the procedure outlined in Ref. \cite{solverdeg} to extract the rapidities. 

\begin{figure}[ht]
\includegraphics[width=0.48\textwidth]{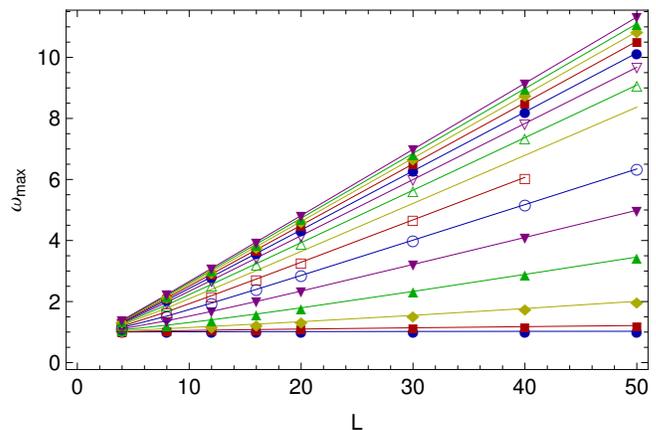}
  \caption{Largest eigenvalue of the matrix $C$ as a function of $L$ for  different coupling strengths, $g=0.1,\dots,1.5$.}
\label{fig:fig2}
\end{figure}

\begin{figure}[ht]
\includegraphics[width=0.48\textwidth]{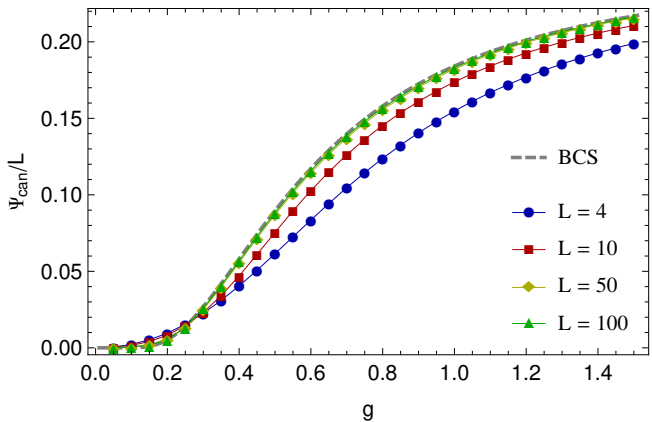}
  \caption{Order parameter $\Psi_{\ms{can}}$ as a function of coupling strength for different system sizes. The dashed line represents the BCS result in the thermodynamic limit.}
\label{fig:fig3}
\end{figure}

\begin{figure}[ht]
\includegraphics[width=0.48\textwidth]{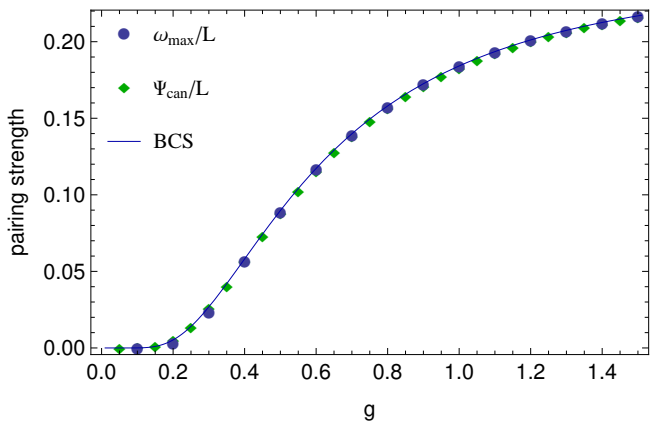}
  \caption{Pairing strength, as measured by $\Psi_{\ms{can}}/L$ (diamonds) and $\lim_{L\rightarrow\infty}(\omega_{\mbox{\scriptsize{max}}}/L)$ (dots). The full line retraces the BCS result.}
\label{fig:fig4}
\end{figure}

Results for the largest eigenvalue $\o_{\ms{max}}$ of $C$ are depicted in Fig. \ref{fig:fig2} as functions of $L$ for various coupling strengths. A linear behaviour is clearly observed already for modest system sizes with slopes that agree perfectly well with BCS theory. Fig. \ref{fig:fig3} shows the exact results for the quantity $\Psi_{\ms{can}}/L$, which also converges rapidly towards the BCS prediction as $L$ increases. Therefore the relation $\o_{\ms{max}}=\Psi_{\ms{can}}$ is also found to hold for the exact ground state and $\o_{\ms{max}}$ can be used interchangeably as order parameter. The results are summarized in Fig. \ref{fig:fig4} where the exact values for $\o_{\ms{max}}/L$ and $\Psi_{\ms{can}}/L$ at large $L$ are seen to agree both with each other and with BCS theory. We conclude that the natural canonical order parameter can be defined either by Eq. (\ref{op}) or as the largest eigenvalue of the reduced density matrix $C$. Both quantities are faithfully predicted by BCS theory.

\section{Correlation functions}
\label{sec5}

We have shown above that the ground state energy $E_0$, the largest eigenvalue 
$\omega_{\ms{max}}$ of the reduced density matrix $C$ and the order parameter $\Psi_{\ms{can}}$ are correctly predicted by BCS theory as the system size $L$ tends to infinity. The same is true for the level occupancy $\langle n_\nu\rangle$, i.e. for the diagonal elements of $C$. But what about non-diagonal matrix elements of $C$, i.e. correlation functions 
$\langle \Psi_0|\bd_\m b_\n|\Psi_0\rangle$ with $\m\neq\n$? To answer this question we have studied the special case where $\m$ is the lowest unoccupied ``molecular orbital" (LUMO) and $\n$ the highest occupied level (HOMO), i.e. $\varepsilon_\mu=-\varepsilon_\nu=(2L)^{-1}$. 

\begin{figure}[ht]
\includegraphics[width=0.48\textwidth]{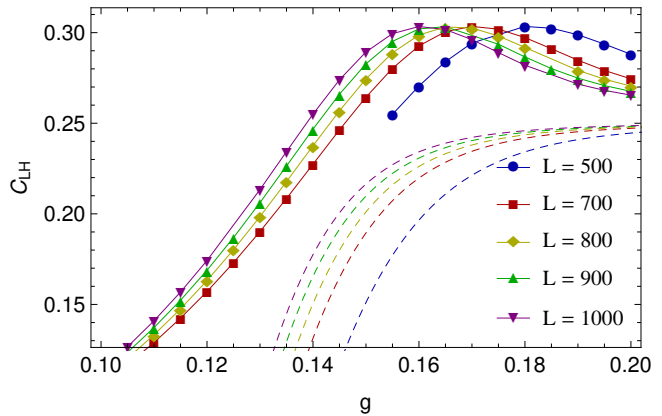}
\caption{HOMO-LUMO pair-pair correlation function. Symbols represent the exact solution, while the dashed lines stand for the BCS result (with $L$ increasing from right to left).}
\label{fig:fig5}
\end{figure}

\begin{figure}[ht]
\includegraphics[width=0.48\textwidth]{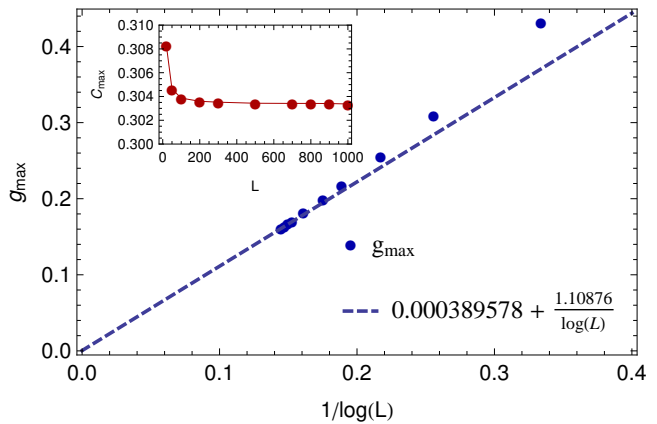}
\caption{Maximum $C_{\mbox{\scriptsize max}}$ and its location $g_{\mbox{\scriptsize max}}$ for the HOMO-LUMO pair-pair correlation function.}
\label{fig:fig6}
\end{figure}

For the conventional BCS ground state we find $C_{\ms{LH}}^{\ms{BCS}}=(L\D)^2/[1+(2L\D)^2]$, where $\D$ represents the gap parameter for $L$ levels ($M=\frac{L}{2}$ pairs). $C_{\ms{LH}}^{\ms{BCS}}$ vanishes for $g<g_{\ms{c}}(L)$ and is finite for $g>g_{\ms{c}}(L)$. Results for the exact ground state are shown in Fig. \ref{fig:fig5} and compared to the BCS predictions. There is good agreement for large coupling strengths but, in contrast to BCS, slightly above $g_c(L)$ there is a peak which does not decrease with increasing system size. We have extracted both the peak values $C_{\ms{max}}$ and the locations of the maxima $g_{\ms{max}}$ by fitting the numerical data with polynomials. The results shown in Fig. \ref{fig:fig6} confirm that the maximum saturates at a value of about $0.30$ and its location $g_{\ms{max}}$ tends to a very small value, consistent with $0$. While BCS theory predicts a simple step at $g=0$ of size $\frac{1}{4}$ in the thermodynamic limit, our analysis indicates that the exact solution exhibits a larger step at $g=0$, followed by a smooth decrease towards the asymptotic strong-coupling limit $\frac{1}{4}$.

\begin{figure}[ht]
\includegraphics[width=0.48\textwidth]{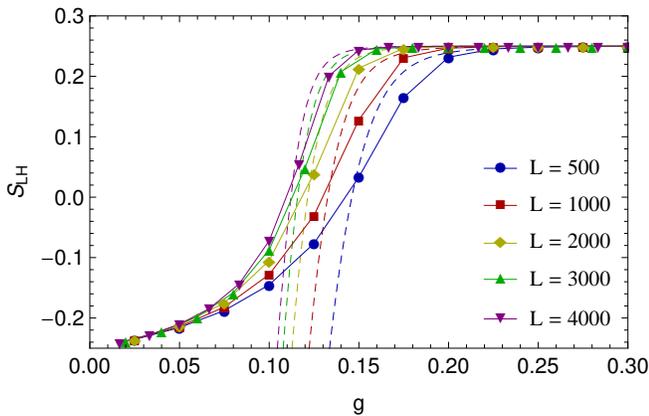}
\caption{$S_{\ms{LH}}$ as a function of coupling strength for different system sizes.}
\label{fig:fig7}
\end{figure}

As a second example we consider the pseudospin-pseudospin correlation function
\bea
S_{\m\n}=\langle\Psi_0|\vec{s}_\m\cdot \vec{s}_\n|\Psi_0\rangle\, ,
\eea
for which BCS theory predicts 
\bea
S_{\m\n}^{\ms{BCS}}=\frac{1}{2}\d_{\m\n}+\frac{\e_\m\e_\n+\D^2}{4E_\m E_\n}.
\eea
For the HOMO-LUMO levels we get 
$S_{\ms{LH}}^{\ms{BCS}}=\frac{1}{4}[(2L\Delta)^2-1]/[(2L\Delta)^2+1]$, which tends to $\frac{1}{4}$ for $L\ra\infty$.
It is straightforward to calculate $S_{\m\n}$ for the exact ground state using the $R$-operators defined by Eq. (\ref{eq:rop}). The ground state is an eigenstate of these operators with eigenvalues
\bea
r_\n=\frac{g}{4L}\sum_{\m,\m\ne\n}\frac{1}{\e_\m-\e_\n}+\L_\n,
\eea
where $\L_\n$ is given by Eq. (\ref{eq:substituted}). Using the Hellman-Feynman theorem for $r_\n=\langle \Psi_0|R_\n|\Psi_0\rangle$ we find (for $\m\ne\n$)
\bea
\frac{\partial r_\n}{\partial\e_\m}&=-\frac{g}{4L}\frac{1}{(\e_\m-\e_\n)^2}+\frac{\partial \L_\n}{\partial \e_\m}=\frac{\partial}{\partial \e_\m}\langle \Psi_0|R_\n|\Psi_0\rangle\nn\\
&=-\frac{g}{L}\frac{1}{(\e_\m-\e_\n)^2}\langle \Psi_0|\vec{s}_\m\cdot\vec{s}_\n|\Psi_0\rangle
\eea
and therefore
\bea
S_{\m\n}=\frac{1}{4}-\frac{L}{g}(\e_\m-\e_\n)^2\frac{\partial\L_\n}{\partial\e_\m}.
\eea
The pseudospin-pseudospin correlation function depends only on the quantities $\L_\n$ (and not explicitly on the rapidities $\l_i$) and therefore can be readily evaluated for large system sizes. Fig. \ref{fig:fig7} shows results for the HOMO-LUMO correlation function $S_{\ms{LH}}$, in comparison with the BCS prediction. Clearly the exact results for $S_{\ms{LH}}$ approach the BCS prediction for $g> g_{\ms{c}}(L)$, and the curves merge more and more rapidly as $L$ increases. These results indicate that the pseudospin-pseudospin correlation function is reproduced exactly by BCS theory for any value of $g$ in the thermodynamic limit.

\begin{figure}[ht]
\includegraphics[width=0.48\textwidth]{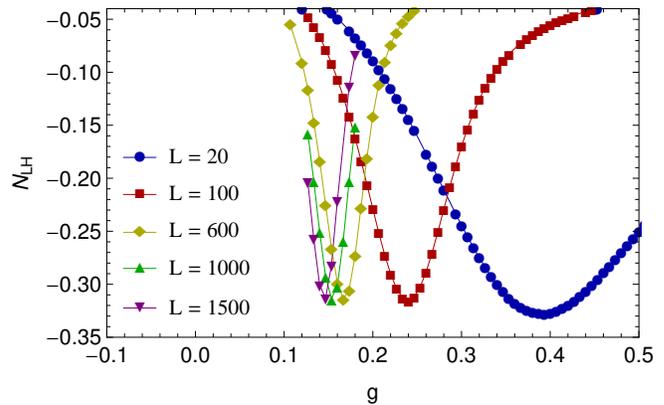}
\caption{HOMO-LUMO occupancy fluctuations for different system sizes.}
\label{fig:fig8}
\end{figure}

\begin{figure}[ht]
\includegraphics[width=0.48\textwidth]{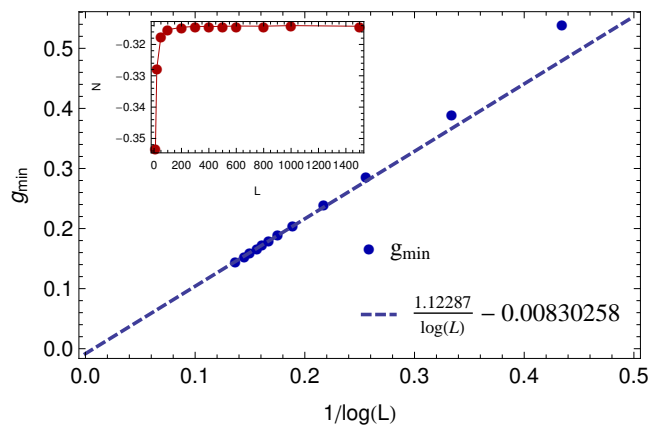}
\caption{Minima $N_{\ms{min}}$ of HOMO-LUMO occupancy fluctuations (inset) and corresponding locations $g_{\ms{min}}$ as functions of $(\log L)^{-1}$.}
\label{fig:fig9}
\end{figure}

\begin{figure}[ht]
\includegraphics[width=0.48\textwidth]{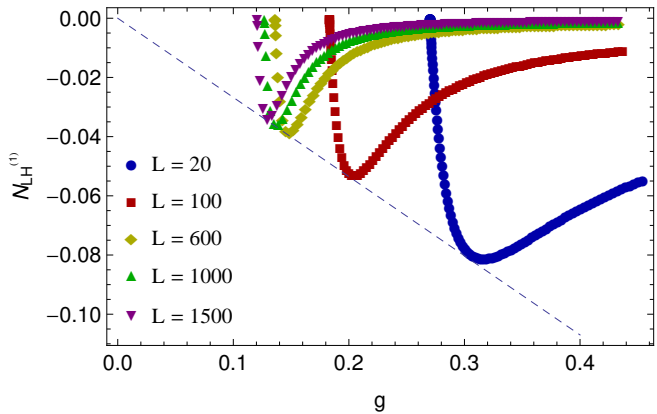}
\caption{HOMO-LUMO occupancy fluctuations according to first order perturbation theory about the BCS ground state. The dashed line represents the asymptotic behavior 
$N_{\ms{min}}^{(1)}\sim -0.2675 g_{\ms{min}}$.}
\label{fig:fig10}
\end{figure}

The pair-pair correlation function (\ref{eq:dens_matr}) can be written in pseudospin language as
\begin{align}
C_{\mu\nu}&=\langle\Psi_0\vert (s_{\m x}s_{\n x}+s_{\m y}s_{\n y})\vert\Psi_0\rangle\nonumber\\
&=S_{\m\n}-\langle\Psi_0\vert s_{\m z}s_{\n z}\vert\Psi_0\rangle\, 
\end{align}
where the (particle-hole) symmetry of $C$ has been used.

According to Eq. (\ref{eq:pseudospin})
$\langle\Psi_0\vert s_{\m z}s_{\n z}\vert\Psi_0\rangle$ measures correlations between level occupancies. It is illuminating to consider fluctuations of these correlations, i.e.
\bea
N_{\m\n}:=\langle\Psi_0\vert 
(n_{\m}-\langle n_{\m}\rangle)(n_{\n}-\langle n_{\n}\rangle)\vert\Psi_0\rangle\, .
\label{eq:fluc_occ}
\eea

$N_{\mu\nu}$ vanishes according to BCS, but, in view of our previous findings for $C_{\m\n}$ and $S_{\m\n}$ it should differ from BCS and thus remain finite for the exact ground state, even in the thermodynamic limit. This is indeed found by our numerical analysis, as shown in Fig. \ref{fig:fig8} for the HOMO-LUMO occupancy fluctuations, which exhibit a pronounced minimum located slightly above $g_c(L)$. While the location of the minimum 
$g_{\ms{min}}$ moves to the left as the system size increases its value $N_{\ms{min}}$ remains essentially constant. This is clearly seen in Fig. {\ref{fig:fig9}} where $N_{\ms{min}}$ and  
$g_{\ms{min}}$ are plotted against $L$ and $1/\log L$, respectively. For large $L$ $g_{\ms{min}}\sim 1.12/\log L$, in close agreement with the corresponding behavior of the pair-pair correlations (Fig. \ref{fig:fig6}). In order to understand better this behavior we have performed a perturbative analysis about the BCS mean-field ground state. Details are given in Appendix \ref{app:A}. We also find clear minima, as shown in Fig. \ref{fig:fig10}, but in contrast to the exact analysis not only the locations of the minima decrease with $L$ but also their values. This can be seen explicitly from the first-order result 
\be
N_{\mbox{\scriptsize LH}}^{(1)}=-g\frac{(2L\Delta)^2[1+2(L\Delta)^2]}{[1+(2L\Delta)^2]^{\frac{5}{2}}}. 
\ee
For large values of $L$ the dominant $g$-dependence of this function is through the gap parameter $\Delta$, with a minimum for $(2L\Delta)^2=\sqrt{5}-1$. Moreover the minimum value is simply proportional to its location, 
$N_{\ms{min}}\approx -0.2675 g_{\ms{min}}$. For large $L$ and small $g$ we can use the 
relation $\Delta\approx e^{-1/g}$ and obtain
\bea
N_{\ms{min}}^{(1)}\approx\frac{-0.2675}{\log L+0.5871795}\, .
\eea
We see that in first-order perturbation theory the minimum value of these fluctuations tends logarithmically to zero as a function of system size, while it remains constant in a full treatment. This suggests that first-order perturbation theory becomes more and more unreliable when approaching criticality, i.e. for  $L\rightarrow\infty$, $g\approx g_c(L)\rightarrow 0$. We expect therefore that in the thermodynamic limit the critical behavior exhibits non-perturbative corrections beyond the BCS mean-field behavior.
\section{Fidelity susceptibility}
\label{sec6}
A sensitive probe of fluctuations is the fidelity susceptibility $\chi_F$, which is often used in the context of quantum phase transitions \cite{zanardi,gu} and can also characterize crossover phenomena  \cite{khanpieri,gut}. For the reduced BCS Hamiltonian $\chi_F$ may be defined as 
\bea
\chi_F(g)=-\frac{2}{L}\lim_{\delta g\rightarrow 0}\frac{\log F(g,\delta g)}{(\delta g)^2},
\label{fidelity}
\eea 
where the fidelity $F(g,\delta g)$ is equal to the overlap 
$\langle\Psi_0(g)\vert\Psi_0(g+\delta g)\rangle$ between ground states associated with infinitesimally close coupling constants. $\chi_F(g)$ can be represented with respect to the eigenstates $\vert\Psi_n(g)\rangle$ of the Hamiltonian with coupling constant $g$, by using ordinary perturbation theory in powers of 
$\delta g$. One finds
\bea
\chi_F(g)=\frac{1}{L}\sum_{n\neq 0}\sum_{\mu,\nu,\mu\neq\nu}\frac{\vert\langle\Psi_0(g)\vert b_\nu^\dag b_\mu\vert\Psi_n(g)\rangle\vert^2}{[E_0(g)-E_n(g)]^2}.
\label{fid_pert}
\eea
Therefore, in contrast to the correlation functions studied in Section \ref{sec5}, the fidelity susceptibility probes all the eigenstates of the Hamiltonian and not only the ground state.
The energy eigenvalues $E_n(g)$ converge to the BCS values for $L\rightarrow\infty$, but this may not be true for all the matrix elements in the numerator. 

\begin{figure}
\includegraphics[width=0.48\textwidth]{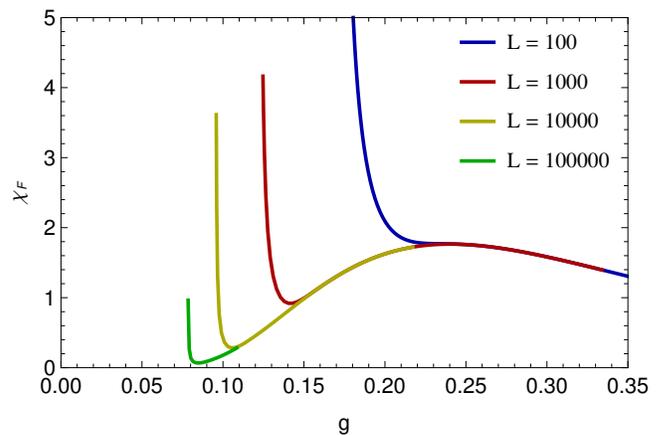}
\caption{Fidelity susceptibility of BCS  theory for $L=10^2, 10^3, 10^4, 10^5$ (from right to left).}
\label{fig:fig11}
\end{figure}

\begin{figure}[ht]
\includegraphics[width=0.48\textwidth]{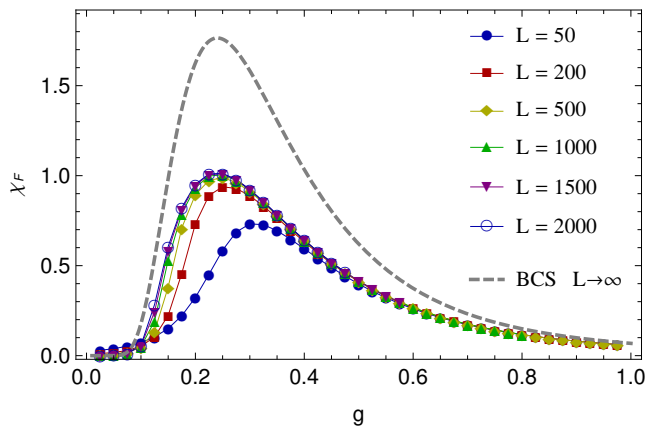}
\caption{Fidelity susceptibility as a function of coupling strength $g$. Symbols on the lower curves represent numerical results for the exact ground state and various values of $L$. The dashed line stands for the BCS result for $L\ra\infty$ .}
\label{fig:fig12}
\end{figure}

\begin{figure}[ht]
\includegraphics[width=0.48\textwidth]{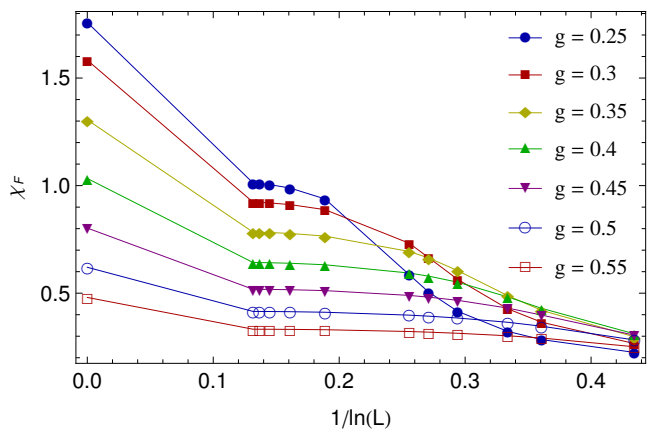}
\caption{Fidelity susceptibility as a function of $1/\ln{L}$ for different coupling strengths $g$. The BCS results are given at $1/\ln{L}=0$.}
\label{fig:fig13}
\end{figure}

In conventional BCS theory the fidelity susceptibility can be obtained analytically. For the case studied here we obtain 
\bea
\chi_F^{\ms{BCS}}(g)=\left(\frac{d\Delta}{dg}\right)^2\frac{1}{4L}\sum_\nu\frac{\varepsilon_\nu^2}{E_\nu^4}\, .
\eea
This function vanishes for $g<g_c(L)$ and diverges if $g$ approaches $g_c(L)$ from above, as shown in Fig. \ref{fig:fig11}. The size of the singularity at $g_c(L)$ decreases with increasing $L$ and disappears for $L\rightarrow\infty$, where $\chi_F$ is given by
\bea
\chi_F^{\ms{BCS}}(g)=\frac{\Delta}{4g^4}\left[(1+4\Delta^2)\arctan\frac{1}{2\Delta}-2\Delta\right]\, ,
\eea
with the asymptotic behavior
\bea
\chi_F^{\ms{BCS}}(g)\sim\frac{\pi}{8g^2}e^{-\frac{1}{g}}
\eea
for $g\rightarrow 0$. There is no divergence at the critical point in the thermodynamic limit, instead there is a broad maximum for $g\approx 0.26$, representing a crossover from the small $g$ to the large $g$ behavior. 

In the Bethe ansatz framework the fidelity $F(g,\delta g)$ is given by the determinant of an 
$L\times L$ matrix \cite{alexdet}, from which the fidelity susceptibility is calculated using 
Eq. (\ref{fidelity}). Fig. \ref{fig:fig12} shows the exact results obtained in this way for different system sizes in comparison with the BCS result for $L\rightarrow\infty$. We observe a rapid convergence to a limiting curve for $g>g_{\ms{c}}(L)$. This is confirmed by a detailed data analysis, illustrated in Fig. \ref{fig:fig13}.  Clearly the exact fidelity susceptibility levels off at a different value than the BCS prediction. The difference is largest around $g\approx 0.26$ (more than $50\%$), which is also the region where both BCS and exact results exhibit a maximum. 

 One may wonder whether the discrepancy between BCS and exact results for the fidelity susceptibility disappears if, instead of the conventional BCS ansatz, we use the number-projected state $|\Psi_{\ms{BCS}}^{(M)}\rangle$. To deal with the number-projected BCS ansatz we have adapted a recursive scheme, used previously for calculating the ground state energy \cite{dukelsky}. Details are given in Appendix \ref{app:B}. The results shown in Fig. \ref{fig:fig14}  indicate a clear convergence between conventional and projected BCS states. Therefore the discrepancy between BCS and the exact solution cannot be removed by replacing the conventional BCS ansatz by the number-projected state.

\begin{figure}[ht]
\includegraphics[width=0.48\textwidth]{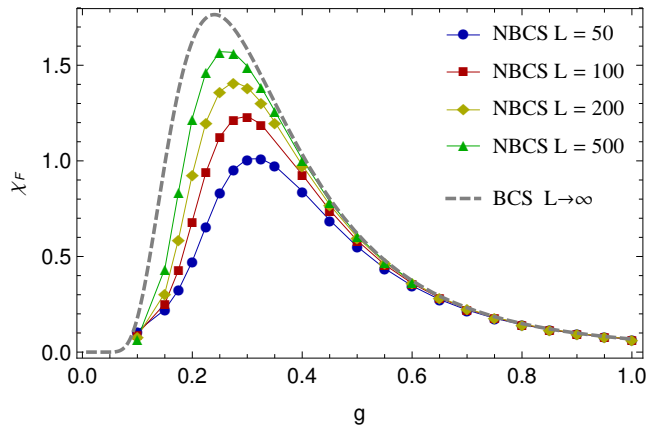}
\caption{Fidelity susceptibility as a function of coupling strength $g$ for the number-projected BCS ground state and different system sizes. The dashed line represents the conventional BCS result for $L\ra\infty$.}
\label{fig:fig14}
\end{figure}
\section{Discussion}
\label{sec7}
In this paper we have studied the exact ground state of the reduced BCS Hamiltonian
for large system sizes. We have confirmed that both the ground state energy $E_0$ and the level occupancies $\langle n_\nu\rangle$ agree with the BCS predictions in the thermodynamic limit. A canonical order parameter $\Psi_{\ms{can}}$, defined either through the concept of ODLRO or by Eq. (\ref{op}), was also found to tend asymptotically to the BCS value. The same turned out to be true for a pseudospin-pseudospin correlation function. These results support the conventional wisdom according to which the mean-field treatment of the reduced BCS Hamiltonian is exact in the thermodynamic limit. However, we did find counterexamples for which the exact results differ from those of BCS theory in this limit, namely the fidelity susceptibility, a pair-pair correlation function and inter-level occupancy fluctuations. In this sense the BCS ground state is not exact in all respects. 

The large $L$ behavior of the two correlation functions for which discrepancies have been found suggests that fluctuations produce non-perturbative corrections to mean-field critical behavior for $L\rightarrow\infty,~g\rightarrow 0$. It would be very interesting to explore this possibility in more depth, for instance using field-theoretical techniques. Another direction of research could be the calculation of dynamic response or correlation functions, for which the discrepancies may be stronger and at the same time easier to measure than the quantities considered here. 

We have limited ourselves to $s$-wave pairing, but an integrable model with $p+ip$ pairing \cite{dunning} could also be analyzed in a similar way. We do not expect any significant differences because for $p+ip$ pairing the density of states around the Fermi energy is completely gapped, as for $s$-wave pairing. An interesting case where the discrepancy between BCS and exact ground state could be more severe than in these integrable systems would be a pairing Hamiltonian where the gap parameter has nodes on the Fermi surface (as for $d$-wave pairing in two dimensions). 

\section*{Acknowledgements}

We are grateful to Vladimir Gritsev for both his continuous interest in our work and helpful suggestions. At the initial stage of our studies we have profited from the experience of Bruno Gut, who presented related issues in his PhD thesis. We also thank Alexandre Faribault, Emil Yuzbashyan and Stijn de Baerdemacker for stimulating discussions, as well as Willi Zwerger for insightful comments. This work has been supported by the Swiss National Science Foundation.
\appendix
\section{Perturbation theory}
\label{app:A}
Bogoliubov's version of BCS theory is based on the mean-field Hamiltonian
\be
H_m=\sum_{\nu\sigma}\varepsilon_\nu c_{\nu\sigma}^\dag c_{\nu\sigma}
-\Delta\sum_\nu (c_{\nu\uparrow}^\dag c_{\nu\downarrow}^\dag
+c_{\nu\downarrow}c_{\nu\uparrow})\, ,
\ee
which is diagonalized by a unitary transformation from fermion operators $c_{\nu\sigma}$ to quasiparticle operators $\gamma_{\nu\sigma}$, i.e.
\be
H_m=\Omega_0+\sum_{\nu\sigma} E_\nu\gamma_{\nu\sigma}^\dag\gamma_{\nu\sigma}\, ,
\ee
where $E_\nu=\sqrt{\varepsilon_\nu^2+\Delta^2}$ and $\Omega_0=-\sum_\nu E_\nu$\, .
The mean-field ground state $\vert\Psi_m\rangle$ is the vacuum of quasiparticles,  
$\gamma_{\nu\sigma}\vert \Psi_m\rangle=0$ for all $\nu,\sigma$. The expectation value of 
the Richardson Hamiltonian (\ref{reddbcs}) with respect to $\vert\Psi_m\rangle$ gives the mean-field ground state energy
\be
E_m=\sum_\nu\left(\varepsilon_\nu-\frac{\varepsilon_\nu^2}{E_\nu}+\frac{g\Delta^2}{4LE_\nu^2}\right)
-\frac{g}{L}\left(\sum_\nu\frac{\Delta}{2E_\nu}\right)^2\, .
\ee
The term of order $1/L$ in the first sum is negligible in the thermodynamic limit, but for finite $L$ it has a small effect on the critical value $g_c(L)$, above which there is a finite gap $\Delta$, and on the value of $\Delta$ for $g>g_c(L)$. Without this term the minimization of $E_0$ with respect to $\Delta$ yields the familiar gap equation
\be
1=\frac{g}{2L}\sum_\nu\frac{1}{E_\nu}\, ,
\label{eq:gap}
\ee
which will be used in the following. We have verified that this approximation has negligible effects for large values of $L$.

We now set up a perturbative expansion around the mean-field solution. To do so, we introduce the ``bare'' Hamiltonian
\be
H_0=H_m+E_m-\Omega_0\, ,
\ee
and the perturbation
\be
H'=&-\frac{g}{L}\sum_{\mu,\nu,\mu\neq\nu}c_{\mu\uparrow}^\dag c_{\mu\downarrow}^\dag
c_{\nu\downarrow}c_{\nu\uparrow}+\Delta\sum_\nu(c_{\nu\uparrow}^\dag c_{\nu\downarrow}^\dag+
c_{\nu\downarrow}c_{\nu\uparrow})\nn\\
&-E_m+\Omega_0\, .
\ee
The Richardson Hamiltonian is then simply given by
\be
H=H_0+H'
\ee
and we may expand in powers of $H'$. We note that the first order correction to the ground state energy vanishes, $\langle\Psi_m\vert H'\vert\Psi_m\rangle=0$. The first order correction to the ground state is found to be
\be
\vert\Psi'\rangle=&
-\frac{g\Delta}{4L}\sum_\nu\frac{\varepsilon_\nu}{E_\nu^3}\beta_\nu^\dag\vert\Psi_m\rangle\nn\\
&-\frac{g}{4L}\sum_{\mu,\nu,\mu<\nu}\frac{(E_\mu E_\nu-\varepsilon_\mu\varepsilon_\nu)}
{E_\mu E_\nu(E_\mu+E_\nu)}\beta_\mu^\dag\beta_\nu^\dag\vert\Psi_m\rangle\, .
\ee
where $\beta_\nu^\dag:=\gamma_{\nu\uparrow}^\dag \gamma_{\nu\downarrow}^\dag$ creates pairs of quasiparticles.

It is straightforward to calculate correlation functions to first order in $H'$. For the pair-pair correlation function (\ref{eq:dens_matr}) we obtain to first order in $H'$ 
\begin{align}
C_{\mu\nu}=&\langle\Psi_m\vert b_\mu^\dag b_\nu\vert\Psi_m\rangle
+\langle\Psi_m\vert( b_\mu^\dag b_\nu +b_\nu^\dag b_\mu)\vert\Psi'\rangle\nonumber\\
=&\frac{\Delta^2}{4E_\mu E_\nu}+
\frac{g}{8LE_\mu E_\nu}\left\{-\Delta^2
\left(\frac{\varepsilon_\mu^2}{E_\mu^3}+\frac{\varepsilon_\nu^2}{E_\nu^3}\right)\right.\nn\\
&\left.+\frac{(E_\mu E_\nu-\varepsilon_\mu\varepsilon_\nu)^2}{E_\mu E_\nu(E_\mu+E_\nu)}\right\}\, .
\end{align}
We consider now the special case where the two levels correspond, respectively, to the ``highest occupied molecular orbital'' (HOMO) and to the ``lowest unoccupied molecular orbital'' (LUMO), i.e. $\varepsilon_\mu=-\varepsilon_\nu=1/(2L)$. We find 
\be
C_{LH}=\frac{(L\Delta)^2}{1+4(L\Delta)^2}
+\frac{g}{2}\frac{1+4(L\Delta)^4}{[1+4(L\Delta)^2]^{\frac{5}{2}}}\, .
\ee
Proceeding in the same way for the occupancy fluctuations (\ref{eq:fluc_occ}) we find to first order in $H'$
\be
N_{\mu\nu}=-\frac{g\Delta^2}{2L}
\frac{E_\mu E_\nu-\varepsilon_\mu\varepsilon_\nu}{E_\mu^2E_\nu^2(E_\mu E_\nu)}\, .
\ee
For the special case of HOMO-LUMO levels we get
\be
N_{\mbox{\scriptsize LH}}=-g\frac{(2L\Delta)^2[1+2(L\Delta)^2]}{[1+(2L\Delta)^2]^{\frac{5}{2}}}. 
\ee
\section{Recursive method for the number-projected BCS state}
\label{app:B}

The BCS pair operator
\be
B^\dag =\sum_\nu \left(\frac{E_\nu-\varepsilon_\nu}{E_\nu+\varepsilon_\nu}\right)^{\frac{1}{2}} b_\nu^\dag
\ee
generates the number-projected BCS ground state
\be
\vert\Psi^{(M)}\rangle=(B^\dag)^M\vert 0\rangle\, .
\ee
Both the norm of the ground state and the expectation value of the Hamiltonian can be calculated recursively \cite{dukelsky}. We have used the recursive scheme for determining the gap parameter for given system sizes $L=2M$ and coupling strengths $g$. We show now how to adapt this method for calculating the fidelity
\be
F(g,g')=\frac{\langle\Psi_m^{(M)}\vert\Psi_m^{'(M)}\rangle}
{\sqrt{\langle\Psi_m^{(M)}\vert\Psi_m^{(M)}\rangle\, \langle\Psi_m^{'(M)}\vert\Psi_m^{'(M)}}\rangle}\, ,
\ee
where $\vert\Psi^{'(M)}\rangle$ is the ground state for the coupling strength $g'$.

The action of the operators $b_\nu,n_\nu$ on $\vert\Psi^{(M)}\rangle$ is given by
\begin{align}
b_\nu\vert\Psi^{(M)}\rangle&=Mf_\nu\vert\Psi^{(M-1)}\rangle
-M(M-1)f_\nu^2b_\nu^\dag\vert\Psi^{(M-2)}\rangle\, ,\nonumber\\
n_\nu\vert\Psi^{(M)}\rangle&=2Mf_\nu b_\nu^\dag\vert\Psi^{(M-1)}\rangle\, ,
\end{align}
leading to a recursion relation for the norm
\be
Z^{(M)}:=\langle\Psi^{(M)}\vert\Psi^{(M)}\rangle\, ,
\ee
namely
\be
Z^{(M)}=MZ^{(M-1)}\sum_\nu f_\nu^2-M(M-1)\sum_\nu f_\nu^3S_\nu^{(M-1)}\, ,
\ee
where
\be
S_\nu^{(M)}:=\langle \Psi^{(M)}\vert b_\nu^\dag\vert\Psi^{(M-1)}\rangle
\ee
is calculated through
\be
S_\nu^{(M)}=Mf_\nu Z^{(M-1)}-M(M-1)f_\nu^2S_\nu^{(M-1)}\, .
\ee
Corresponding relations hold for $Z^{'(M)}$ and $S_\nu^{'(M)}$, while the overlap 
\be
V^{(M)}:=\langle\Psi^{(M)}\vert\Psi^{'(M)}\rangle
\ee
is obtained recursively as
\be
V^{(M)}=&MV^{(M-1)}\sum_\nu f_\nu f'_\nu\nn\\
&-M(M-1)\sum_\nu f_\nu^2 f'_\nu W_\nu^{(M-1)}\, ,
\ee
where
\be
W_\nu^{(M)}:=\langle \Psi^{'(M)}\vert b_\nu^\dag\vert\Psi^{(M-1)}\rangle\, .
\ee
One also needs the quantity
\be
Y_\nu^{(M)}:=\langle\Psi^{(M)}\vert b_\nu^\dag\vert\Psi^{'(M-1)}\rangle\, .
\ee
The system is closed by the recursion relations for $W_\nu^{(M)}$ and $Y_\nu^{(M)}$,
\begin{align}
W_\nu^{(M)}=Mf'_\nu V^{(M-1)}-M(M-1)f_\nu^{'2}Y_\nu^{(M-1)}\, ,\nonumber\\
Y_\nu^{(M)}=Mf_\nu V^{(M-1)}-M(M-1)f_\nu^2W_\nu^{(M-1)}\, ,
\end{align}
together with the initial conditions
\begin{align}
Z^{(1)}&=\sum_\nu f_\nu^2\, , \quad Z^{'(1)}=\sum_\nu f_\nu^{'2}\, , \quad
V^{(1)}=\sum_\nu f_\nu f'_\nu\, .\nonumber\\
S_\nu^{(1)}&=Y_\nu^{(1)}=f_\nu\, ,~~~~
S_\nu^{'(1)}=W_\nu^{(1)}=f'_\nu\, .
\end{align}

\bibliography{refpaper}

\begin{thebibliography}{36}%
\makeatletter
\providecommand \@ifxundefined [1]{%
 \@ifx{#1\undefined}
}%
\providecommand \@ifnum [1]{%
 \ifnum #1\expandafter \@firstoftwo
 \else \expandafter \@secondoftwo
 \fi
}%
\providecommand \@ifx [1]{%
 \ifx #1\expandafter \@firstoftwo
 \else \expandafter \@secondoftwo
 \fi
}%
\providecommand \natexlab [1]{#1}%
\providecommand \enquote  [1]{``#1''}%
\providecommand \bibnamefont  [1]{#1}%
\providecommand \bibfnamefont [1]{#1}%
\providecommand \citenamefont [1]{#1}%
\providecommand \href@noop [0]{\@secondoftwo}%
\providecommand \href [0]{\begingroup \@sanitize@url \@href}%
\providecommand \@href[1]{\@@startlink{#1}\@@href}%
\providecommand \@@href[1]{\endgroup#1\@@endlink}%
\providecommand \@sanitize@url [0]{\catcode `\\12\catcode `\$12\catcode
  `\&12\catcode `\#12\catcode `\^12\catcode `\_12\catcode `\%12\relax}%
\providecommand \@@startlink[1]{}%
\providecommand \@@endlink[0]{}%
\providecommand \url  [0]{\begingroup\@sanitize@url \@url }%
\providecommand \@url [1]{\endgroup\@href {#1}{\urlprefix }}%
\providecommand \urlprefix  [0]{URL }%
\providecommand \Eprint [0]{\href }%
\providecommand \doibase [0]{http://dx.doi.org/}%
\providecommand \selectlanguage [0]{\@gobble}%
\providecommand \bibinfo  [0]{\@secondoftwo}%
\providecommand \bibfield  [0]{\@secondoftwo}%
\providecommand \translation [1]{[#1]}%
\providecommand \BibitemOpen [0]{}%
\providecommand \bibitemStop [0]{}%
\providecommand \bibitemNoStop [0]{.\EOS\space}%
\providecommand \EOS [0]{\spacefactor3000\relax}%
\providecommand \BibitemShut  [1]{\csname bibitem#1\endcsname}%
\let\auto@bib@innerbib\@empty
\bibitem [{\citenamefont {Bardeen}\ \emph {et~al.}(1957)\citenamefont
  {Bardeen}, \citenamefont {Cooper},\ and\ \citenamefont {Schrieffer}}]{bcs1}%
  \BibitemOpen
  \bibfield  {author} {\bibinfo {author} {\bibfnamefont {J.}~\bibnamefont
  {Bardeen}}, \bibinfo {author} {\bibfnamefont {L.~N.}\ \bibnamefont {Cooper}},
  \ and\ \bibinfo {author} {\bibfnamefont {J.~R.}\ \bibnamefont {Schrieffer}},\
  }\href@noop {} {\bibfield  {journal} {\bibinfo  {journal} {Phys. Rev.}\
  }\textbf {\bibinfo {volume} {108}},\ \bibinfo {pages} {1175} (\bibinfo {year}
  {1957})}\BibitemShut {NoStop}%
\bibitem [{\citenamefont {Leggett}(1975)}]{leggett}%
  \BibitemOpen
  \bibfield  {author} {\bibinfo {author} {\bibfnamefont {A.~J.}\ \bibnamefont
  {Leggett}},\ }\href@noop {} {\bibfield  {journal} {\bibinfo  {journal} {Rev.
  Mod. Phys.}\ }\textbf {\bibinfo {volume} {47}},\ \bibinfo {pages} {331}
  (\bibinfo {year} {1975})}\BibitemShut {NoStop}%
\bibitem [{\citenamefont {Giorgini}\ \emph {et~al.}(2008)\citenamefont
  {Giorgini}, \citenamefont {Pitaevskii},\ and\ \citenamefont
  {Stringari}}]{giorgini}%
  \BibitemOpen
  \bibfield  {author} {\bibinfo {author} {\bibfnamefont {S.}~\bibnamefont
  {Giorgini}}, \bibinfo {author} {\bibfnamefont {L.~P.}\ \bibnamefont
  {Pitaevskii}}, \ and\ \bibinfo {author} {\bibfnamefont {S.}~\bibnamefont
  {Stringari}},\ }\href@noop {} {\bibfield  {journal} {\bibinfo  {journal}
  {Rev. Mod. Phys.}\ }\textbf {\bibinfo {volume} {80}},\ \bibinfo {pages}
  {1215} (\bibinfo {year} {2008})}\BibitemShut {NoStop}%
\bibitem [{\citenamefont {Bloch}\ \emph {et~al.}(2008)\citenamefont {Bloch},
  \citenamefont {Dalibard},\ and\ \citenamefont {Zwerger}}]{bloch}%
  \BibitemOpen
  \bibfield  {author} {\bibinfo {author} {\bibfnamefont {I.}~\bibnamefont
  {Bloch}}, \bibinfo {author} {\bibfnamefont {J.}~\bibnamefont {Dalibard}}, \
  and\ \bibinfo {author} {\bibfnamefont {W.}~\bibnamefont {Zwerger}},\
  }\href@noop {} {\bibfield  {journal} {\bibinfo  {journal} {Rev. Mod. Phys.}\
  }\textbf {\bibinfo {volume} {80}},\ \bibinfo {pages} {885} (\bibinfo {year}
  {2008})}\BibitemShut {NoStop}%
\bibitem [{\citenamefont {Broglia}\ and\ \citenamefont
  {Zelevinsky}()}]{broglia}%
  \BibitemOpen
  \bibfield  {author} {\bibinfo {author} {\bibfnamefont {R.~A.}\ \bibnamefont
  {Broglia}}\ and\ \bibinfo {author} {\bibfnamefont {V.}~\bibnamefont
  {Zelevinsky}},\ }\href@noop {} {\bibinfo  {journal} {{\it 50 Years of Nuclear
  BCS}, World Scientific 2013}\ }\BibitemShut {NoStop}%
\bibitem [{\citenamefont {Rajagopal}\ and\ \citenamefont
  {Wilczek}()}]{wilczek}%
  \BibitemOpen
\bibfield  {journal} {  }\bibfield  {author} {\bibinfo {author} {\bibfnamefont
  {K.}~\bibnamefont {Rajagopal}}\ and\ \bibinfo {author} {\bibfnamefont
  {F.}~\bibnamefont {Wilczek}},\ }\href@noop {} {\bibinfo  {journal} {Handbook
  of QCD, Chapter 35, World Scientific 2001}\ }\BibitemShut {NoStop}%
\bibitem [{\citenamefont {Anderson}(1958)}]{anderson}%
  \BibitemOpen
\bibfield  {journal} {  }\bibfield  {author} {\bibinfo {author} {\bibfnamefont
  {P.~W.}\ \bibnamefont {Anderson}},\ }\href@noop {} {\bibfield  {journal}
  {\bibinfo  {journal} {Phys. Rev.}\ }\textbf {\bibinfo {volume} {112}},\
  \bibinfo {pages} {1900} (\bibinfo {year} {1958})}\BibitemShut {NoStop}%
\bibitem [{\citenamefont {Bogoliubov}\ \emph {et~al.}(1961)\citenamefont
  {Bogoliubov}, \citenamefont {Zubarev},\ and\ \citenamefont
  {Tserkovnikov}}]{bogoliubov}%
  \BibitemOpen
  \bibfield  {author} {\bibinfo {author} {\bibfnamefont {N.~N.}\ \bibnamefont
  {Bogoliubov}}, \bibinfo {author} {\bibfnamefont {D.~N.}\ \bibnamefont
  {Zubarev}}, \ and\ \bibinfo {author} {\bibfnamefont {I.~A.}\ \bibnamefont
  {Tserkovnikov}},\ }\href@noop {} {\bibfield  {journal} {\bibinfo  {journal}
  {Sov. Phys. JETP}\ }\textbf {\bibinfo {volume} {12}},\ \bibinfo {pages} {88}
  (\bibinfo {year} {1961})}\BibitemShut {NoStop}%
\bibitem [{\citenamefont {M\"uhlschlegel}(1962)}]{muehlschlegel}%
  \BibitemOpen
  \bibfield  {author} {\bibinfo {author} {\bibfnamefont {B.}~\bibnamefont
  {M\"uhlschlegel}},\ }\href@noop {} {\bibfield  {journal} {\bibinfo  {journal}
  {J. Math. Phys.}\ }\textbf {\bibinfo {volume} {3}},\ \bibinfo {pages} {522}
  (\bibinfo {year} {1962})}\BibitemShut {NoStop}%
\bibitem [{\citenamefont {Bursill}\ and\ \citenamefont
  {Thompson}(1993)}]{bursill}%
  \BibitemOpen
  \bibfield  {author} {\bibinfo {author} {\bibfnamefont {R.~J.}\ \bibnamefont
  {Bursill}}\ and\ \bibinfo {author} {\bibfnamefont {C.~J.}\ \bibnamefont
  {Thompson}},\ }\href@noop {} {\bibfield  {journal} {\bibinfo  {journal} {J.
  Phys. A: Math. and Gen.}\ }\textbf {\bibinfo {volume} {26}},\ \bibinfo
  {pages} {769} (\bibinfo {year} {1993})}\BibitemShut {NoStop}%
\bibitem [{\citenamefont {Mattis}\ and\ \citenamefont
  {Lieb}(1961)}]{mattislieb}%
  \BibitemOpen
  \bibfield  {author} {\bibinfo {author} {\bibfnamefont {D.}~\bibnamefont
  {Mattis}}\ and\ \bibinfo {author} {\bibfnamefont {E.}~\bibnamefont {Lieb}},\
  }\href@noop {} {\bibfield  {journal} {\bibinfo  {journal} {J. Math. Phys.}\
  }\textbf {\bibinfo {volume} {2}},\ \bibinfo {pages} {602} (\bibinfo {year}
  {1961})}\BibitemShut {NoStop}%
\bibitem [{\citenamefont {Richardson}(1963)}]{richardson1}%
  \BibitemOpen
  \bibfield  {author} {\bibinfo {author} {\bibfnamefont {R.}~\bibnamefont
  {Richardson}},\ }\href@noop {} {\bibfield  {journal} {\bibinfo  {journal}
  {Phys. Lett.}\ }\textbf {\bibinfo {volume} {3}},\ \bibinfo {pages} {277 }
  (\bibinfo {year} {1963})}\BibitemShut {NoStop}%
\bibitem [{\citenamefont {Richardson}\ and\ \citenamefont
  {Sherman}(1964)}]{richardson2}%
  \BibitemOpen
  \bibfield  {author} {\bibinfo {author} {\bibfnamefont {R.}~\bibnamefont
  {Richardson}}\ and\ \bibinfo {author} {\bibfnamefont {N.}~\bibnamefont
  {Sherman}},\ }\href@noop {} {\bibfield  {journal} {\bibinfo  {journal} {Nucl.
  Phys.}\ }\textbf {\bibinfo {volume} {52}},\ \bibinfo {pages} {221 } (\bibinfo
  {year} {1964})}\BibitemShut {NoStop}%
\bibitem [{\citenamefont {Yang}(1962)}]{yangodlro}%
  \BibitemOpen
  \bibfield  {author} {\bibinfo {author} {\bibfnamefont {C.~N.}\ \bibnamefont
  {Yang}},\ }\href@noop {} {\bibfield  {journal} {\bibinfo  {journal} {Rev.
  Mod. Phys.}\ }\textbf {\bibinfo {volume} {34}},\ \bibinfo {pages} {694}
  (\bibinfo {year} {1962})}\BibitemShut {NoStop}%
\bibitem [{\citenamefont {Gaudin}(1976)}]{gaudindiago}%
  \BibitemOpen
  \bibfield  {author} {\bibinfo {author} {\bibfnamefont {M.}~\bibnamefont
  {Gaudin}},\ }\href@noop {} {\bibfield  {journal} {\bibinfo  {journal} {J.
  Phys. France}\ }\textbf {\bibinfo {volume} {37}},\ \bibinfo {pages} {1087}
  (\bibinfo {year} {1976})}\BibitemShut {NoStop}%
\bibitem [{\citenamefont {Cambiaggio}\ \emph {et~al.}(1997)\citenamefont
  {Cambiaggio}, \citenamefont {Rivas},\ and\ \citenamefont
  {Saraceno}}]{cambiaggio}%
  \BibitemOpen
  \bibfield  {author} {\bibinfo {author} {\bibfnamefont {M.}~\bibnamefont
  {Cambiaggio}}, \bibinfo {author} {\bibfnamefont {A.}~\bibnamefont {Rivas}}, \
  and\ \bibinfo {author} {\bibfnamefont {M.}~\bibnamefont {Saraceno}},\
  }\href@noop {} {\bibfield  {journal} {\bibinfo  {journal} {Nucl. Phys. A}\
  }\textbf {\bibinfo {volume} {624}},\ \bibinfo {pages} {157 } (\bibinfo {year}
  {1997})}\BibitemShut {NoStop}%
\bibitem [{\citenamefont {Faribault}\ \emph {et~al.}(2011)\citenamefont
  {Faribault}, \citenamefont {El~Araby}, \citenamefont {Str\"ater},\ and\
  \citenamefont {Gritsev}}]{solver}%
  \BibitemOpen
  \bibfield  {author} {\bibinfo {author} {\bibfnamefont {A.}~\bibnamefont
  {Faribault}}, \bibinfo {author} {\bibfnamefont {O.}~\bibnamefont {El~Araby}},
  \bibinfo {author} {\bibfnamefont {C.}~\bibnamefont {Str\"ater}}, \ and\
  \bibinfo {author} {\bibfnamefont {V.}~\bibnamefont {Gritsev}},\ }\href@noop
  {} {\bibfield  {journal} {\bibinfo  {journal} {Phys. Rev. B}\ }\textbf
  {\bibinfo {volume} {83}},\ \bibinfo {pages} {235124} (\bibinfo {year}
  {2011})}\BibitemShut {NoStop}%
\bibitem [{\citenamefont {Gaudin}()}]{gaudinmodexres}%
  \BibitemOpen
  \bibfield  {author} {\bibinfo {author} {\bibfnamefont {M.}~\bibnamefont
  {Gaudin}},\ }\href@noop {} {\bibinfo  {journal} {{\it Mod{\`e}les exactement
  r{\'e}solus}, Les {\'E}ditions de Physique, 1995}\ }\BibitemShut {NoStop}%
\bibitem [{\citenamefont {Richardson}(1977)}]{richardson3}%
  \BibitemOpen
\bibfield  {journal} {  }\bibfield  {author} {\bibinfo {author} {\bibfnamefont
  {R.}~\bibnamefont {Richardson}},\ }\href@noop {} {\bibfield  {journal}
  {\bibinfo  {journal} {J. Math. Phys.}\ }\textbf {\bibinfo {volume} {18}},\
  \bibinfo {pages} {1802} (\bibinfo {year} {1977})}\BibitemShut {NoStop}%
\bibitem [{\citenamefont {Rom\'an}\ \emph {et~al.}(2002)\citenamefont
  {Rom\'an}, \citenamefont {Sierra},\ and\ \citenamefont {Dukelsky}}]{sierra}%
  \BibitemOpen
  \bibfield  {author} {\bibinfo {author} {\bibfnamefont {J.}~\bibnamefont
  {Rom\'an}}, \bibinfo {author} {\bibfnamefont {G.}~\bibnamefont {Sierra}}, \
  and\ \bibinfo {author} {\bibfnamefont {J.}~\bibnamefont {Dukelsky}},\
  }\href@noop {} {\bibfield  {journal} {\bibinfo  {journal} {Nucl. Phys. B}\
  }\textbf {\bibinfo {volume} {634}},\ \bibinfo {pages} {483 } (\bibinfo {year}
  {2002})}\BibitemShut {NoStop}%
\bibitem [{\citenamefont {Yuzbashyan}\ \emph {et~al.}(2003)\citenamefont
  {Yuzbashyan}, \citenamefont {Baytin},\ and\ \citenamefont
  {Altshuler}}]{emil}%
  \BibitemOpen
  \bibfield  {author} {\bibinfo {author} {\bibfnamefont {E.~A.}\ \bibnamefont
  {Yuzbashyan}}, \bibinfo {author} {\bibfnamefont {A.~A.}\ \bibnamefont
  {Baytin}}, \ and\ \bibinfo {author} {\bibfnamefont {B.~L.}\ \bibnamefont
  {Altshuler}},\ }\href@noop {} {\bibfield  {journal} {\bibinfo  {journal}
  {Phys. Rev. B}\ }\textbf {\bibinfo {volume} {68}},\ \bibinfo {pages} {214509}
  (\bibinfo {year} {2003})}\BibitemShut {NoStop}%
\bibitem [{\citenamefont {Combescot}\ \emph {et~al.}(2013)\citenamefont
  {Combescot}, \citenamefont {Pogosov},\ and\ \citenamefont
  {Betbeder-Matibet}}]{combescot}%
  \BibitemOpen
  \bibfield  {author} {\bibinfo {author} {\bibfnamefont {M.}~\bibnamefont
  {Combescot}}, \bibinfo {author} {\bibfnamefont {W.~V.}\ \bibnamefont
  {Pogosov}}, \ and\ \bibinfo {author} {\bibfnamefont {O.}~\bibnamefont
  {Betbeder-Matibet}},\ }\href@noop {} {\bibfield  {journal} {\bibinfo
  {journal} {Physica C}\ }\textbf {\bibinfo {volume} {485}},\ \bibinfo {pages}
  {47} (\bibinfo {year} {2013})}\BibitemShut {NoStop}%
\bibitem [{\citenamefont {Braun}\ and\ \citenamefont {von
  Delft}(1998)}]{braun}%
  \BibitemOpen
  \bibfield  {author} {\bibinfo {author} {\bibfnamefont {F.}~\bibnamefont
  {Braun}}\ and\ \bibinfo {author} {\bibfnamefont {J.}~\bibnamefont {von
  Delft}},\ }\href@noop {} {\bibfield  {journal} {\bibinfo  {journal} {Phys.
  Rev. Lett.}\ }\textbf {\bibinfo {volume} {81}},\ \bibinfo {pages} {4712}
  (\bibinfo {year} {1998})}\BibitemShut {NoStop}%
\bibitem [{\citenamefont {Dukelsky}\ and\ \citenamefont
  {Sierra}(2000)}]{dukelsky}%
  \BibitemOpen
  \bibfield  {author} {\bibinfo {author} {\bibfnamefont {J.}~\bibnamefont
  {Dukelsky}}\ and\ \bibinfo {author} {\bibfnamefont {G.}~\bibnamefont
  {Sierra}},\ }\href@noop {} {\bibfield  {journal} {\bibinfo  {journal} {Phys.
  Rev. B}\ }\textbf {\bibinfo {volume} {61}},\ \bibinfo {pages} {12302}
  (\bibinfo {year} {2000})}\BibitemShut {NoStop}%
\bibitem [{\citenamefont {Gor'kov}(1959)}]{gorkov}%
  \BibitemOpen
  \bibfield  {author} {\bibinfo {author} {\bibfnamefont {L.~P.}\ \bibnamefont
  {Gor'kov}},\ }\href@noop {} {\bibfield  {journal} {\bibinfo  {journal}
  {JETP}\ }\textbf {\bibinfo {volume} {9}},\ \bibinfo {pages} {1364} (\bibinfo
  {year} {1959})}\BibitemShut {NoStop}%
\bibitem [{\citenamefont {von Delft}\ \emph {et~al.}(1996)\citenamefont {von
  Delft}, \citenamefont {Zaikin}, \citenamefont {Golubev},\ and\ \citenamefont
  {Tichy}}]{vonDelft}%
  \BibitemOpen
  \bibfield  {author} {\bibinfo {author} {\bibfnamefont {J.}~\bibnamefont {von
  Delft}}, \bibinfo {author} {\bibfnamefont {A.~D.}\ \bibnamefont {Zaikin}},
  \bibinfo {author} {\bibfnamefont {D.~S.}\ \bibnamefont {Golubev}}, \ and\
  \bibinfo {author} {\bibfnamefont {W.}~\bibnamefont {Tichy}},\ }\href@noop {}
  {\bibfield  {journal} {\bibinfo  {journal} {Phys. Rev. Lett.}\ }\textbf
  {\bibinfo {volume} {77}},\ \bibinfo {pages} {3189} (\bibinfo {year}
  {1996})}\BibitemShut {NoStop}%
\bibitem [{\citenamefont {Tian}\ \emph {et~al.}(2000)\citenamefont {Tian},
  \citenamefont {Tang},\ and\ \citenamefont {Chen}}]{Tian}%
  \BibitemOpen
  \bibfield  {author} {\bibinfo {author} {\bibfnamefont {G.-S.}\ \bibnamefont
  {Tian}}, \bibinfo {author} {\bibfnamefont {L.-H.}\ \bibnamefont {Tang}}, \
  and\ \bibinfo {author} {\bibfnamefont {Q.-H.}\ \bibnamefont {Chen}},\
  }\href@noop {} {\bibfield  {journal} {\bibinfo  {journal} {Europhys. Lett}\
  }\textbf {\bibinfo {volume} {50}},\ \bibinfo {pages} {361} (\bibinfo {year}
  {2000})}\BibitemShut {NoStop}%
\bibitem [{\citenamefont {Faribault}\ \emph {et~al.}(2008)\citenamefont
  {Faribault}, \citenamefont {Calabrese},\ and\ \citenamefont
  {Caux}}]{alexcorr}%
  \BibitemOpen
  \bibfield  {author} {\bibinfo {author} {\bibfnamefont {A.}~\bibnamefont
  {Faribault}}, \bibinfo {author} {\bibfnamefont {P.}~\bibnamefont
  {Calabrese}}, \ and\ \bibinfo {author} {\bibfnamefont {J.-S.}\ \bibnamefont
  {Caux}},\ }\href@noop {} {\bibfield  {journal} {\bibinfo  {journal} {Phys.
  Rev. B}\ }\textbf {\bibinfo {volume} {77}},\ \bibinfo {pages} {064503}
  (\bibinfo {year} {2008})}\BibitemShut {NoStop}%
\bibitem [{\citenamefont {von Delft}\ and\ \citenamefont
  {Ralph}(2001)}]{vondelft1}%
  \BibitemOpen
  \bibfield  {author} {\bibinfo {author} {\bibfnamefont {J.}~\bibnamefont {von
  Delft}}\ and\ \bibinfo {author} {\bibfnamefont {D.~C.}\ \bibnamefont
  {Ralph}},\ }\href@noop {} {\bibfield  {journal} {\bibinfo  {journal} {Phys.
  Rep.}\ }\textbf {\bibinfo {volume} {345}},\ \bibinfo {pages} {61} (\bibinfo
  {year} {2001})}\BibitemShut {NoStop}%
\bibitem [{\citenamefont {El~Araby}\ \emph {et~al.}(2012)\citenamefont
  {El~Araby}, \citenamefont {Gritsev},\ and\ \citenamefont
  {Faribault}}]{solverdeg}%
  \BibitemOpen
  \bibfield  {author} {\bibinfo {author} {\bibfnamefont {O.}~\bibnamefont
  {El~Araby}}, \bibinfo {author} {\bibfnamefont {V.}~\bibnamefont {Gritsev}}, \
  and\ \bibinfo {author} {\bibfnamefont {A.}~\bibnamefont {Faribault}},\
  }\href@noop {} {\bibfield  {journal} {\bibinfo  {journal} {Phys. Rev. B}\
  }\textbf {\bibinfo {volume} {85}},\ \bibinfo {pages} {115130} (\bibinfo
  {year} {2012})}\BibitemShut {NoStop}%
\bibitem [{\citenamefont {Zanardi}\ and\ \citenamefont
  {Paunkovi\'c}(2006)}]{zanardi}%
  \BibitemOpen
  \bibfield  {author} {\bibinfo {author} {\bibfnamefont {P.}~\bibnamefont
  {Zanardi}}\ and\ \bibinfo {author} {\bibfnamefont {N.}~\bibnamefont
  {Paunkovi\'c}},\ }\href@noop {} {\bibfield  {journal} {\bibinfo  {journal}
  {Phys. Rev. E}\ }\textbf {\bibinfo {volume} {74}},\ \bibinfo {pages} {031123}
  (\bibinfo {year} {2006})}\BibitemShut {NoStop}%
\bibitem [{\citenamefont {Gu}(2010)}]{gu}%
  \BibitemOpen
  \bibfield  {author} {\bibinfo {author} {\bibfnamefont {S.-J.}\ \bibnamefont
  {Gu}},\ }\href@noop {} {\bibfield  {journal} {\bibinfo  {journal} {Int. J.
  Mod. Phys. B}\ }\textbf {\bibinfo {volume} {24}},\ \bibinfo {pages} {4371}
  (\bibinfo {year} {2010})}\BibitemShut {NoStop}%
\bibitem [{\citenamefont {Khan}\ and\ \citenamefont {Pieri}(2009)}]{khanpieri}%
  \BibitemOpen
  \bibfield  {author} {\bibinfo {author} {\bibfnamefont {A.}~\bibnamefont
  {Khan}}\ and\ \bibinfo {author} {\bibfnamefont {P.}~\bibnamefont {Pieri}},\
  }\href@noop {} {\bibfield  {journal} {\bibinfo  {journal} {Phys. Rev. A}\
  }\textbf {\bibinfo {volume} {80}},\ \bibinfo {pages} {012303} (\bibinfo
  {year} {2009})}\BibitemShut {NoStop}%
\bibitem [{\citenamefont {Gut}(2009)}]{gut}%
  \BibitemOpen
  \bibfield  {author} {\bibinfo {author} {\bibfnamefont {B.~J.}\ \bibnamefont
  {Gut}},\ }\href@noop {} {Ph.D. thesis},\ \bibinfo  {school} {University of
  Fribourg} (\bibinfo {year} {2009})\BibitemShut {NoStop}%
\bibitem [{\citenamefont {Faribault}\ and\ \citenamefont
  {Schuricht}(2012)}]{alexdet}%
  \BibitemOpen
  \bibfield  {author} {\bibinfo {author} {\bibfnamefont {A.}~\bibnamefont
  {Faribault}}\ and\ \bibinfo {author} {\bibfnamefont {D.}~\bibnamefont
  {Schuricht}},\ }\href@noop {} {\bibfield  {journal} {\bibinfo  {journal} {J.
  Phys. A: Math. and Theor.}\ }\textbf {\bibinfo {volume} {45}},\ \bibinfo
  {pages} {485202} (\bibinfo {year} {2012})}\BibitemShut {NoStop}%
\bibitem [{\citenamefont {Dunning}\ \emph {et~al.}(2010)\citenamefont
  {Dunning}, \citenamefont {Ibanez}, \citenamefont {Links}, \citenamefont
  {Sierra},\ and\ \citenamefont {Zhao}}]{dunning}%
  \BibitemOpen
  \bibfield  {author} {\bibinfo {author} {\bibfnamefont {C.}~\bibnamefont
  {Dunning}}, \bibinfo {author} {\bibfnamefont {M.}~\bibnamefont {Ibanez}},
  \bibinfo {author} {\bibfnamefont {J.}~\bibnamefont {Links}}, \bibinfo
  {author} {\bibfnamefont {G.}~\bibnamefont {Sierra}}, \ and\ \bibinfo {author}
  {\bibfnamefont {S.-Y.}\ \bibnamefont {Zhao}},\ }\href@noop {} {\bibfield
  {journal} {\bibinfo  {journal} {J. Stat. Mech.}\ }\textbf {\bibinfo {volume}
  {P08025}} (\bibinfo {year} {2010})}\BibitemShut {NoStop}%
\end{thebibliography}%
\end{document}